\documentclass[11pt,oneside]{article}
\usepackage{a4wide}
\usepackage{mathrsfs}
\usepackage{latexsym,bm}
\usepackage{graphicx}
\usepackage{indentfirst}
\usepackage{slashed}
\usepackage{amsmath}
\usepackage{amssymb}
\usepackage{color}
\usepackage{hyperref}
\usepackage{epsfig}
\usepackage[titletoc]{appendix}
\usepackage{multirow}%
\usepackage{rotating}
\usepackage{cite}
\usepackage{ulem} 
\usepackage[top=2.9cm,bottom=2.5cm,left=2.8cm,right=3cm]{geometry}
\usepackage{tabularx}

\newcommand{\email}[1]{\footnote{{\em } \texttt{#1}}}

\newcommand{\sigcp}{\Sigma_c^{+}}
\newcommand{\sigcpp}{\Sigma_c^{++}}
\newcommand{\dpn}{\bar{D}^{0}}
\newcommand{\dpm}{D^{-}}
\newcommand{\dvn}{\bar{D}^{*0}}
\newcommand{\dvm}{D^{*-}}
\newcommand{\jpsi}{J/\psi}
\newcommand{\pca}{P_c(4312)}
\newcommand{\pcb}{P_c(4440)}
\newcommand{\pcc}{P_c(4457)}
\newcommand{\al}{&\!\!\!\!}

\begin{document}
\thispagestyle{empty}
\title{
\Large \bf  Anatomy of the newly observed hidden-charm pentaquark states: $P_c(4312)$, $P_c(4440)$ and $P_c(4457)$  }
\author{\small Zhi-Hui Guo$^{a}$\email{zhguo@hebtu.edu.cn}, \,  J.~A.~Oller$^{b}$\email{oller@um.es}  \\[0.5em]
{ \small\it ${}^a$  Department of Physics and Hebei Advanced Thin Films Laboratory, } \\
{\small\it Hebei Normal University,  Shijiazhuang 050024, China}\\[0.3em]
{\small {\it $^b$Departamento de F\'{\i}sica. Universidad de Murcia. E-30071 Murcia. Spain}}
}
\date{}

%

\maketitle
\begin{abstract}
We study the newly reported hidden-charm pentaquark candidates $P_c(4312)$, $P_c(4440)$ and $P_c(4457)$ from the LHCb Collaboration, in the framework of the effective-range expansion and resonance compositeness relations. The scattering lengths and effective ranges from the $S$-wave $\Sigma_c\bar{D}$ and $\Sigma_c\bar{D}^*$ scattering are calculated by using the experimental results of the masses and widths of the $P_c(4312)$, $P_c(4440)$ and $P_c(4457)$. Then we calculate the couplings between the $J/\psi p,\,\Sigma_c\bar{D}$ channels and the pentaquark candidate $P_c(4312)$, with which we further estimate the probabilities of finding the $J/\psi p$ and $\Sigma_c\bar{D}$ components inside $P_c(4312)$. The partial decay widths and compositeness coefficients are calculated for the $P_c(4440)$ and $P_c(4457)$ states by including the $J/\psi p$ and $\Sigma_c\bar{D}^*$ channels. Similar studies are also carried out for the three $P_c$ states by including the $\Lambda_c\bar{D}^{*}$ and $\Sigma_c\bar{D}^{(*)}$ channels. 
\end{abstract}

\section{Introduction}

The first discovery of the hidden-charm pentaquark states $P_c(4380)$ and $P_c(4450)$~\cite{Aaij:2015tga} has triggered a plethora of in-depth theoretical studies~\cite{ref.review}. Very recently, the LHCb Collaboration has reported updated results on the pentaquark states based on the combinations of the Run~1 + Run~2 data~\cite{Aaij:2019vzc}. The first notable finding from the updated measurements is that a new hidden-charm pentaquark state $\pca$ is observed with the mass $4311.9\pm0.7^{+6.8}_{-0.6}$~MeV and the width $9.8\pm2.7^{+3.7}_{-4.5}$~MeV. The second notable and intriguing observation is that the previous single state $P_c(4450)$ is superseded by two nearby states $\pcb$ and $\pcc$, with their masses $4440.3\pm1.3^{+4.1}_{-4.7}$~MeV and $4457.3\pm0.6^{+4.1}_{-1.7}$~MeV,  respectively, and their widths $20.6\pm4.9^{+8.7}_{-10.1}$~MeV and $6.4\pm2.0^{+5.7}_{-1.9}$~MeV, respectively. The previous peak around the  $P_c(4380)$ state now becomes less clear and its existence needs to be confirmed further by the experimental analysis. The new measurements have already attracted attention from many  groups~\cite{Chen:2019bip,Chen:2019asm,Guo:2019fdo,Liu:2019tjn,He:2019ify,Huang:2019jlf,Xiao:2019mvs,Shimizu:2019ptd,Ali:2019npk,Cao:2019kst}.

All of the three new states are observed in the $\jpsi p$ invariant mass distributions from the $\Lambda_b \to \jpsi\,p\,K^-$ decay. One of the common features of the newly measured pentaquark states is that they all have small widths. Another important common feature is that they lie quite close to the thresholds of two underlying hadrons. In the following discussions we take a conservative way to estimate the experimental values of the masses and widths for the $\pca,\pcb$ and $\pcc$~\cite{Aaij:2019vzc}.
To be more specific, we  
take the larger values in magnitude  
of the upper or lower limits for the systematic uncertainties, and  
add them quadratically to the statistical ones as the total uncertainties.
The resulting masses and widths are summarized in the second and third columns of  Table~\ref{tab.ar}. 
The differences between the mass of the $\pca$ and the $\sigcp\dpn$ and $\sigcpp\dpm$ thresholds are $5.8\pm 6.8$~MeV and $11.7\pm 6.8$~MeV, respectively.  The mass of $\pcb$ lie $19.5\pm 4.9$~MeV and $23.9\pm 4.9$~MeV below the $\sigcp\dvn$ and $\sigcpp\dvm$ thresholds, respectively. For the $\pcc$, the differences between its mass and the $\sigcp\dvn$ and $\sigcpp\dvm$ thresholds are $2.5\pm 4.1$~MeV and $6.9\pm 4.1$~MeV, respectively. 
Taking into account the uncertainties of the experimental measurements of the $\pca$,
we notice that its mass can be either below or above the $\sigcp\dpn$ threshold,
but it is always below the $\sigcpp\dpm$ threshold. 
For the mass of the $\pcc$ a similar situation occurs, so that, within the present experimental uncertainties \cite{Aaij:2019vzc},
its mass can be also either below or above the $\sigcp\dvn$ threshold,
but it is always below the $\sigcpp\dvm$ threshold.
As a result one would expect that the isospin breaking effects could be visible~\cite{Guo:2019fdo}. In order to quantify the possible isospin breaking effects, we shall distinguish the elastic scattering with different thresholds
involving $\sigcp$ or $\sigcpp$ in %
a later study.

In this work, our key aim is to quantify the possibilities of the $\pca$ as the $S$-wave $\Sigma_c\bar{D}$, and the $\pcb$ and $\pcc$ as the $S$-wave $\Sigma_c\bar{D}^{*}$ molecular states. 
The effective range expansion (ERE) approach offers a reliable tool to analyze the dynamics around the threshold energy region. The combinations of the analyticity, unitarity and ERE have been demonstrated to be successful in analyzing the heavy-flavor exotic hadrons near thresholds~\cite{Gao:2018jhk,Kang:2016ezb,Kang:2016jxw,Guo:2016wpy}. Another powerful tool that can help to reveal the inner structures of the hadrons is the Weinberg's compositeness relation~\cite{Weinberg:1962hj}, which is extended to the resonance case in Refs.~\cite{Guo:2015daa,Oller:2017alp,Meissner:2015mza}.
Other forms of generalization %
for other compositeness relation to address the resonances can be also found
in Refs.~\cite{Baru:2003qq,Hanhart:2011jz,Hyodo:2011qc,Aceti:2012dd,Sekihara:2014kya}.
In the current work we combine 
analyticity, unitarity, the ERE and the resonance compositeness relation 
to study the three newly measured pentaquark states.

\section{Effective-range-expansion study of the pentaquark states} 
\label{sec.190401.1}

The ERE approach relies on the power series expansion  of the  
$K$-matrix $V(k)$ at around threshold
\begin{eqnarray}
  \label{190331.1}
V(k) = -\frac{1}{a} + \frac{1}{2} r\,k^2 + O(k^2) \,,
\end{eqnarray}
where $a$ is the scattering length, $r$ denotes the effective range and $k$ stands for the magnitude of three-momentum in the center of mass (CM) frame. 
For a two-particle system with masses $m_1$ and $m_2$, in the non-relativistic limit the three-momentum $k$ is related to the CM energy $E$  through
\begin{eqnarray}
k= \sqrt{2\mu(E-m_{\rm th})}\,,
\end{eqnarray}
with the threshold $m_{\rm th}=m_1 + m_2 $ and the reduced mass $\mu=\frac{m_1m_2}{m_1+m_2}$.

For the $\Sigma_c\bar{D}$ scattering near the $\pca$ and the $\Sigma_c\bar{D}^{*}$ scattering near the $\pcb$ and $\pcc$ energy regions, the magnitudes of the three-momenta of the two-particle systems can range from 0 to 250~MeV, after taking into account the experimental uncertainties of the masses of the $P_c$ states~\cite{Aaij:2019vzc}. 
For the scattering of two heavy-flavor hadrons, it is plausible that the pion exchanges can be treated perturbatively~\cite{Fleming:2007rp,Valderrama:2012jv,Wu:2010jy,Wu:2010vk,Wu:2012md}.
For the heavier vector-resonance exchanges, their contributions can be effectively included via 
contact interactions, since their masses are clearly  larger than the scale of the relevant three-momenta. 
Therefore we take the point of view from the pionless effective field theory,
which only needs to include the local contact interactions \cite{kolck.190419.1}.
 
Under these circumstances only the unitarity/right-hand cut enters and there is no crossed-channel dynamics. 
The elastic $S$-wave scattering amplitude  around  threshold %
that results from Eq.~\eqref{190331.1} (without the crossed-channel cuts) can be written as 
\begin{eqnarray}\label{eq.ti}
T(E)=\frac{1}{-\frac{1}{a}+\frac{1}{2}r\,k^2-i\,k}\,,
\end{eqnarray}
which satisfies the unitarity condition
\begin{equation}
 {\rm Im}\,T(E)^{-1} = - k  \,,\quad (E > m_{\rm th})\,.
\end{equation}

The formula $T(E)$ in Eq.~\eqref{eq.ti} generally works well in the energy region near threshold even when resonances appear, except in the special situation that an underlying Castillejo-Dalitz-Dyson (CDD) pole sits on top of the threshold. In the latter case, one has to explicitly include the CDD pole in Eq.~\eqref{eq.ti} and we refer to Ref.~\cite{Guo:2016wpy} for further details. It is difficult to know whether there is a CDD pole near threshold a priori. Nevertheless in Refs.~\cite{Guo:2016wpy,Kang:2016ezb} it is proved that when a CDD pole approaches to the threshold one has the following behaviors for the scattering length and effective range  
\begin{eqnarray} \label{eq.arcdd}
 a \to -\frac{m_{\rm th}- M_{\rm CDD}}{ g_{\rm CDD}}\,, \quad  r \to -\frac{ g_{\rm CDD}}{\mu(m_{\rm th}- M_{\rm CDD})^2}\,,
\end{eqnarray}
with $M_{\rm CDD}$ the bare CDD pole mass and $g_{\rm CDD}$ the residue. According to Eq.~\eqref{eq.arcdd}, one can infer that there exists a CDD pole near the threshold only for the situations with $|a|\ll 1$~fm and $|r|\gg 1$~fm. In this situation, one should use the formalism developed in Ref.~\cite{Guo:2016wpy} to proceed, instead of Eq.~\eqref{eq.ti}. 

In the present work we 
first blindly use the ERE formalism in Eq.~\eqref{eq.ti}.
If the resulting $a$ and $r$ have natural values of the long-range hadronic scale at $1/m_\pi\sim\,1$~fm,
one could then safely conclude that the formalism in Eq.~\eqref{eq.ti} is applicable in our study
(with no indication of a near-threshold CDD pole). 
We demonstrate below that the resulting values of $a$ and $r$ from the $S$-wave $\Sigma_c\bar{D}$ scattering 
around $\pca$ and  the $S$-wave $\Sigma_c\bar{D}^{*}$ scattering around the $\pcb$ and $\pcc$, 
indeed have typical long-range hadronic scale around $1$~fm. 
Another issue that needs to be clarified is that we implicitly assume a definite 
isospin number for the $P_c$ states (although we do not need to specify it),
otherwise we had to use a coupled-channel scattering formalism in the ERE study.
Regarding the quantum numbers of $J^{P}$, the negative $P$ parity can be uniquely fixed in 
the $S$-wave $\Sigma_c \bar{D}^{(*)}$ scattering.
Similarly, the total angular momentum is $J=1/2$  for $\Sigma_c \bar{D}$ $S$-wave scattering,
while for the analogous $\Sigma_c \bar{D}^{*}$ case there  are two possibilities,  $J=1/2$ or $3/2$, 
which can not be pinned down from our study. 

For a resonance pole, its position $E_R$ is denoted as
\begin{equation}
E_R = M_R -i\Gamma_R/2\,,
\end{equation}
where $M_R$ is the resonance mass and $\Gamma_R$ denotes its width.
The resonance poles lie on the second Riemann sheet (RS) of the scattering amplitude $T_{II}(E)$, which is given by 
\begin{eqnarray}\label{eq.tii}
T_{II}(E)=\frac{1}{-\frac{1}{a}+\frac{1}{2}r\,k^2+i\,k}\,.
\end{eqnarray}
We mention that the convention ${\rm Im}k>0$ should be taken in Eqs.~\eqref{eq.ti} and \eqref{eq.tii}. Given the mass and width of the resonance, we can determine the scattering length $a$ and effective range $r$ by  
requiring %
that $T_{II}(E_R)^{-1}=0$, i.e. 
\begin{eqnarray}\label{eq.getar}
-\frac{1}{a}+\frac{1}{2}r\,k_R^2+i\,k_R=0\,,
\end{eqnarray}
where $k_R$ is the corresponding three-momentum at the pole position
\begin{equation}
k_R= \sqrt{\mu(E_R-m_{\rm th})}\,.  
\end{equation}

By solving Eq.~\eqref{eq.getar}, it is straightforward to determine the values of $a$ and $r$
once the masses and widths of the resonances are given, with the result \cite{Kang:2016ezb}
\begin{align}
a&=-\frac{2k_i}{|k_R|^2}~,\\
r&=-\frac{1}{k_i}~,\nonumber
\end{align}
where $k_r={\rm Re}\, k_R$ and $k_i= {\rm Im}\, k_R$.
As mentioned  
above in the Introduction, we distinguish the different charged states $\sigcp$ and $\sigcpp$,
in order to quantify the isospin breaking effects.
The thresholds of %
the different charged states are explicitly given in the fourth column of Table~\ref{tab.ar}.
The results for the scattering lengths $a$ and effective ranges $r$ with uncertainties are collected in %
the fifth and sixth columns of Table~\ref{tab.ar}, respectively.

\begin{table}[htbp]
\centering
\begin{scriptsize}
\begin{tabular}{ c c c c c c }
\hline\hline
 Resonance & Mass   & Width & Threshold  & $a$     & $r$     \\
           & (MeV)  & (MeV) &  (MeV)    & (fm)    &  (fm)   
\\ \hline
$\pca$ & $4311.9\pm 6.8$ & $9.8\pm 5.2$ & $\sigcp\dpn$~(4317.7)  & $-2.9 \pm 0.8$ & $-1.7 \pm 0.7$  \\
       &   &                            & $\sigcpp\dpm$~(4323.6) & $-2.4 \pm 0.6$ & $-1.2 \pm 0.3$ 
\\ \hline
$\pcb$ & $4440.3\pm 4.9$ & $20.6\pm 11.2$ & $\sigcp\dvn$~(4459.8)  & $-1.7 \pm 0.2$ & $-0.9 \pm 0.1$  \\
       &   &                              & $\sigcpp\dvm$~(4464.2) & $-1.6 \pm 0.2$ & $-0.8 \pm 0.1$ 
\\ \hline
$\pcc$ & $4457.3\pm 4.1$ & $6.4\pm 6.0$ & $\sigcp\dvn$~(4459.8)   & $-3.8 \pm 1.6$ & $-2.3 \pm 1.3$  \\
       &   &                              & $\sigcpp\dvm$~(4464.2) & $-3.0 \pm 0.7$ & $-1.6 \pm 0.4$ 
\\\hline\hline
\end{tabular}
\end{scriptsize}
\caption{\label{tab.ar} The values of %
  the scattering lengths and effective ranges of the $S$-wave amplitudes 
  for different channels.
The uncertainties 
for $a$ and $r$ are determined by %
adding in quadrature the resulting ones from the systematic and statistical errors
of the masses and widths of the $P_c$ states.
The errors of the different thresholds are negligible in comparison with the uncertainties of the masses and widths of
the $P_c$ states.} 
\end{table}

According to the %
values obtained for $a$ and $r$ in Table~\ref{tab.ar},
although we see some discrepancies 
in the central values  
for the channels with different charged states, they are compatible after taking into account the uncertainties. 
It implies that the isospin breaking effects in the three $P_c$ states seem mild and further experimental
reduction of the uncertainties could help to identify the roles of the isospin breaking.

All of the resulting scattering lengths $a$ and effective ranges $r$ in Table~\ref{tab.ar} have  
natural values 
of the order of $1$~fm, indicating that indeed there is no need  
for introducing CDD poles near the thresholds. %
Let us notice that this  outcome  
is consistent with  the application of Eq.~\eqref{eq.ti} in our study.
Furthermore, the natural values of the $a$ and $r$ allow us to qualitatively conclude that the $\pca$
can be described as an $S$-wave $\Sigma_c\bar{D}$ molecular state,
and the $\pcb$ and $\pcc$ are 
$S$-wave $\Sigma_c\bar{D}^*$ composite states.
Nevertheless, in the ERE approach we can not use the prescription in Ref.~\cite{Guo:2015daa} to give 
a quantitative estimate of the probabilities of the $\Sigma_c\bar{D}$ component in the $\pca$ and of
the $\Sigma_c\bar{D}^*$ component in the $\pcb$ and $\pcc$ resonances.
In Ref.~\cite{Guo:2015daa} it has been demonstrated that one can only give the probabilistic
interpretation of the compositeness coefficients when the resonance pole $s_R=E_R^2$ lies in an unphysical 
RS that is directly connected to the physical one in the region $s_k<s<s_{k+1}$, such that $s_k < {\rm Re}s_R< s_{k+1}$, with $s_k$ and $s_{k+1}$ the two nearby thresholds.
In the single-channel scattering case, it requires that the resonance pole %
mass should lie above the threshold in the second RS. 
However in most of cases the pole positions of the $P_c$ states in Table~\ref{tab.ar} are below the thresholds.
This fact refrains us from discussing the probabilities of finding the two-particle components
in the $P_c$ states in the ERE approach. 

In the present formalism we are assuming that the whole width of a resonance is due to the 
corresponding $\Sigma_c \bar{D}^{(*)}$ channel, and the resulting $a$ and $r$ are real.
On general grounds, because of the presence of the inelastic channels below threshold, like the 
$J/\psi p$ one to which these resonances decay, the ERE parameters $a$ and $r$ are complex. 
One possible way to proceed is to include explicitly
the inelastic channels below the $\Sigma_c \bar{D}^{(*)}$ channel, such as the aforementioned $J/\psi p$. 
However, in the coupled-channel scattering case, there would be needed
extra scattering input which is beyond the scope of the present study.
In order to give quantitative information of the inner structures of the $P_c$ states, and take 
into account at least one inelastic channel, 
we proceed the study by relating the compositeness coefficients with the partial decay widths in next section. 

\section{Compositeness relations and the partial widths}

As mentioned previously, we can not access the quantitative information of the constituents inside the $\pca$
in the elastic scattering $\Sigma_c\bar{D}$ and the $\pcb$ and $\pcc$ from the elastic $\Sigma_c\bar{D}^*$ scattering. 
A straightforward extension is to include the additional $\jpsi p$ channel, %
in which invariant-mass distribution the different $P_c$ resonances are actually detected ~\cite{Aaij:2019vzc}.
For the two-channel $\jpsi p$ and $\Sigma_c\bar{D}^{(*)}$ systems, it is natural to assume that the $P_c$ resonances
lie in the second RS, which now allows us to exploit the formalism in Ref.~\cite{Guo:2015daa} to calculate
the probabilities of the two-particle components in the $P_c$.
Analogous study has been carried out for the obsolete $P_c(4450)$ state by including the $\jpsi p$ and $\chi_{c1} p$ channels %
in Ref.~\cite{Meissner:2015mza}.

The essential prescription of Ref.~\cite{Guo:2015daa} to calculate the partial compositeness coefficient
$X_j$ of a resonance $R$ contributed by the $j$th channel is given by 
\begin{eqnarray}\label{eq.xj}
X_j = |g_j|^2  \left| \frac{\partial G_j(s_R)}{\partial s} \right|\,,
\end{eqnarray}
where $g_j$ denotes the coupling between the two-particle state and the resonance $R$, and the one-loop two-point function $G(s)$ is given by
\begin{eqnarray}\label{eq.defg}
G(s)=i\int\frac{{\rm d}^4q}{(2\pi)^4}
\frac{1}{(q^2-m_{1}^2+i\epsilon)[(P-q)^2-m_{2}^2+i\epsilon ]}\ ,\qquad
s\equiv P^2\ \,.
\end{eqnarray} 
This expression 
can be explicitly integrated out by using a once-subtracted dispersion relation or dimensional regularization
(replacing the divergence by a constant), which then reads ~\cite{Oller:1998zr} 
\begin{eqnarray}\label{eq.gfunc}
G(s) \al=\al\frac{1}{16\pi^2}\bigg\{{a}(\mu_g)+\ln\frac{m_{1}^2}{\mu_g^2}
+\frac{s-m_{1}^2+m_{2}^2}{2s}\ln\frac{m_{2}^2}{m_{1}^2}\nonumber\\
\al\al+\frac{\sigma}{2s}\big[\ln(s-m_{2}^2+m_{1}^2+\sigma)-\ln(-s+m_{2}^2-m_{1}^2+\sigma)\nonumber\\
\al\al+\ln(s+m_{2}^2-m_{1}^2+\sigma)-\ln(-s-m_{2}^2+m_{1}^2+\sigma)\big]\bigg\}\ ,
\end{eqnarray}
where 
\begin{equation}
\sigma(s)=\sqrt{[s-(m_1+m_2)^2][s-(m_1-m_2)^2]}\,. 
\end{equation}
The evaluation of $G_j(s)$ for the $j$th channel in Eq.~\eqref{eq.xj}  
requires to use the proper masses $m_1$ and $m_2$ in that channel.
In this equation $\partial G_j(s_R)/\partial s$ 
denotes the partial derivative evaluated at the resonance pole position $s_R=E_R^2=(M_R-i\Gamma_R/2)^2$. 
Notice that $\partial G_j(s_R)/\partial s$ is independent on the subtraction constant $a(\mu_g)$ and the regularization scale $\mu_g$ in Eq.~\eqref{eq.gfunc}. 

In order to fix the two couplings $g_{i=1,2}$, we impose that the decay widths $\Gamma_R$ of the $P_c$ states
are saturated by the two channels $\jpsi p$ and $\Sigma_c\bar{D}^{(*)}$. 
The partial decay width $\Gamma_1$  
to $\jpsi p$ takes the standard form~\cite{Tanabashi:2018oca}
\begin{eqnarray} \label{eq.gamma1}
\Gamma_1=  |g_1|^2  \frac{q(M_R^2)}{8\pi M_R^2}\,,
\end{eqnarray}
where the relativistic three-momentum $q(M_R^2)$ is 
\begin{eqnarray}
 q(M_R^2) =\frac{\sqrt{[M_R^2-(m_{1}+m_{2})^2][M_R^2-(m_{1}-m_{2})^2]}}{2M_R} \,. 
\end{eqnarray}
Since in many cases the masses of the $P_c$ resonances  
are below the thresholds of $\Sigma_c\bar{D}^{(*)}$,
we introduce 
a Lorentzian mass distribution to calculate the partial width $\Gamma_2$ 
to the $\Sigma_c\bar{D}^{(*)}$ channel as  
\begin{eqnarray}\label{eq.gamma2}
\Gamma_2 =  |g_2|^2 \int_{m_{\rm th}}^{M_R+2\Gamma_R} dw \,\frac{q(w^2)}{16\pi^2 \,w^2} \frac{\Gamma_R}{(M_R-w)^2+\Gamma_R^2/4} \,. 
\end{eqnarray}
To restrict the discussion 
to the resonance energy region, we set the upper integration limit at $M_R+2\Gamma_R$ in Eq.~\eqref{eq.gamma2}, as  in Ref.~\cite{Meissner:2015mza}. 
After taking into account Eqs.~\eqref{eq.gamma1} and \eqref{eq.gamma2}, the saturation condition of the $P_c$ decay widths by the $\jpsi p$ and $\Sigma_c\bar{D}^{(*)}$ channels gives
\begin{eqnarray}\label{eq.g12b}
|g_1|^2  \,\frac{q_1(M_R^2)}{8\pi M_R^2} +  |g_2|^2 \int_{m_{\rm th}}^{M_R+2\Gamma_R} dw\, \frac{q_2(w^2)}{16\pi^2 \,w^2} \frac{\Gamma_R}{(M_R-w)^2+\Gamma_R^2/4} &=& \Gamma_R \,,
 \end{eqnarray} 
with $q_1$ and $q_2$ the three-momenta of the $\jpsi p$ and $\Sigma_c\bar{D}^{(*)}$ channels, respectively.

For the resonance poles in the second RS in the coupled-channel $\jpsi p$ and $\Sigma_c\bar{D}^{(*)}$ scattering,
one can identify the compositeness coefficient $X_j$ in Eq.~\eqref{eq.xj} 
as the probability to find the two-particle state from the $j$th channel in the considered resonance.
We mention that within the uncertainties of the masses of the $\pca$ and $\pcc$, %
a tiny portion of their poles lies in the third RS (in which the three-momenta of the two channels flip sign) so that they are continuously connected with the physical RS above the $\Sigma_c\bar{D}^{(*)}$ threshold.
Nevertheless, due to their closeness to the thresholds, their effects can be covered by the large uncertainties
of the $P_c$ masses. Therefore we shall only focus on the poles on the second RS in the following. 

As a clarification remark, let us notice that in Eq.~\eqref{eq.xj} the coupling is taken constant in the range of
masses of the resonance along its Lorentzian mass distribution because of the finite width of the resonance, cf. Eq.~\eqref{eq.gamma2}. 
In this way, there is a smooth transition in the calculation of $X_2$ as the value of the nominal resonance pole mass $M_R$ varies from 
above to below the threshold. This allows us some flexibility in order to bypass the strict  requirement that the resonance mass should lie
above the thresholds of the channels for which $X_j$  is calculated.
However,  in the elastic ERE approach discussed in Sec.~\ref{sec.190401.1},
 the whole width is accounted for only by the channel explicitly taken into account (the second one in the present coupled-channel study),
and the situation is more stringent in this respect \cite{Kang:2016ezb}. 

The total compositeness $X$ is the sum of $X_1$ and $X_2$, with $X_1$ the partial compositeness coefficient of the $\jpsi p$ and $X_2$
the coefficient of $\Sigma_c\bar{D}^{(*)}$. By using Eq.~\eqref{eq.xj}, we can obtain 
\begin{eqnarray}\label{eq.g12a}
 |g_1|^2 \, \left| \frac{\partial G_1^{II}(s_R)}{\partial s} \right| +  |g_2|^2 \, \left| \frac{\partial G_2(s_R)}{\partial s} \right| &=& X \,,  
\end{eqnarray}
where  
$G_1^{II}(s)$ stands for the $G(s)$ function on the second RS and it is related to the expression
in Eq.~\eqref{eq.gfunc} through $G^{II}(s)= G(s)+ i \sigma(s)/(8\pi s)$.

For a given value of the total compositeness $X$ contributed by the $\jpsi p$ and $\Sigma_c\bar{D}^{(*)}$ channels, we can determine the couplings $|g_1|$ and $|g_2|$ by combining Eqs.~\eqref{eq.g12b} and \eqref{eq.g12a}. 
In this way, we can further calculate the partial compositeness coefficients $X_{1,2}$ using Eq.~\eqref{eq.xj}
and the partial decay widths $\Gamma_{1,2}$ via Eqs.~\eqref{eq.gamma1} and \eqref{eq.gamma2}.
In principle the partial widths $\Gamma_2$ consist of combinations of the $\sigcp\bar{D}^{(*)0}$ and $\sigcpp D^{(*)-}$ channels, depending on the isospin  of the pentaquark states $P_c$. 
Nevertheless, we point out that the method employed
is not sensitive to whether we assume a definite isospin for the $P_c$ states or not,
as long as the same masses of the $\Sigma_c \bar{D}^{(*)}$ are taken in Eqs.~\eqref{eq.gamma2} and \eqref{eq.g12b}. 
The reason is because the couplings squared of the different charged states 
simply add together in these equations.
In order to check the isospin breaking effects, we separately solve Eqs.~\eqref{eq.g12b} and \eqref{eq.g12a} by using either the masses of $\sigcp\bar{D}^{0}(\bar{D}^{*0})$ or $\sigcpp D^{-}(D^{*-})$. 

Concerning the value of $X$ in Eq.~\eqref{eq.g12a}, we distinguish three different scenarios.
In the first scenario, we assume that the compositeness of the $P_c$ states  
is completely saturated by the $\jpsi p$  
and 
 $\Sigma_c\bar{D}^{(*)}$channels, that is, we first assume that  $X=1$.
For each $P_c$ state, we separately perform the calculations by using either the masses of $\sigcp\bar{D}^{0}(\bar{D}^{*0})$ or $\sigcpp D^{-}(D^{*-})$.
The resulting values of the couplings $|g_1|$ and $|g_2|$, the partial widths $\Gamma_1$ and $\Gamma_2$,
and the partial compositeness coefficients $X_1$ and $X_2$ are summarized in Table~\ref{tab.x1}. 
The first lesson we learn from Table~\ref{tab.x1} is that the $P_c$ couplings $|g_1|$ to 
the $\jpsi p$ channel are much smaller than the couplings $|g_2|$ to the $\Sigma_c\bar{D}^{(*)}$ channel.
The situation 
for the partial decay widths becomes less clear, since many of them have large uncertainties. 
In all the cases, the overwhelmingly dominant components of the $P_c$ states are found to be the $\Sigma_c\bar{D}^{(*)}$, %
in agreement with our qualitative understanding in Sec.~\ref{sec.190401.1} from the values of $a$ and $r$ given in Table~\ref{tab.ar}.

In 
the next two scenarios, we set the compositeness $X=0.8$ and $0.5$ in Eq.~\eqref{eq.g12a}.
In order not to overload the table, we only show the values obtained by using the masses of $\sigcp$ and $\bar{D}^{(*)0}$ in Table~\ref{tab.xsmall}.
The results by using the masses of $\sigcpp$ and $D^{(*)-}$ are quantitatively similar. 
All the values in Table~\ref{tab.xsmall} show quite similar trends as those in Table~\ref{tab.x1}, with $X_2\gg X_1$.

The previous discussions rely on the assumption that the decay widths of the $P_c$ states are saturated 
by the $\jpsi p$ and $\Sigma_c\bar{D}^{(*)}$ channels.
Other decay patterns are also predicted, such as those in  Refs.~\cite{Shen:2016tzq,Lin:2017mtz},
which suggest that the partial decay widths of the pentaquark states to the $\Lambda_c \bar{D}^{(*)}$ channels 
could be more important than to the $\jpsi p$.
In order to check the robustness of our conclusion, 
we include the $\Lambda_c \bar{D}^{(*)}$ and $\Sigma_c \bar{D}^{(*)}$ channels to perform a similar study. 
To be specific, we give the results in Table~\ref{tab.xlamcdv} by using the masses of 
$\Lambda_c^{+} \bar{D}^{*0}$ and $\Sigma_c^{+} \bar{D}^{*0}$. It is verified that to use 
the masses of other charged states leads to quantitatively similar results. 
Since to replace the $\Lambda_c \bar{D}^{*}$ channel by the $\Lambda_c \bar{D}$ does not 
lead to qualitatively new trends, we do not explicitly show the corresponding results.
Comparing the numbers in Tables~\ref{tab.x1}, \ref{tab.xsmall}
and those in Table~\ref{tab.xlamcdv}, not only the partial decay widths of the two different
sets of dynamical channels are quite similar, but also 
the compositeness coefficients in the different cases are compatible within uncertainties.

\begin{table}[htbp]
\centering
\begin{scriptsize}
\begin{tabular}{ c c c c c c c }
\hline\hline
Resonance & $|g_1|$ & $|g_2|$   & $\Gamma_1$ & $\Gamma_2$  & $X_1$   & $X_2$     \\
          &(GeV)  & (GeV)     & (MeV)      &  (MeV)      &         &     
\\ \hline
$\pca$ & & & & & & \\
$m_{\sigcp} + m_{\dpn}$   & $2.1^{+0.8}_{-2.1}$ & $10.9^{+2.1}_{-2.9}$ & $6.5^{+4.9}_{-6.5}$  & $3.3^{+10.5}_{-3.3}$ & $0.006^{+0.005}_{-0.006}$ &  $0.994^{+0.006}_{-0.005}$ \\
$m_{\sigcpp} + m_{\dpm}$  & $2.5^{+0.6}_{-0.9}$ & $12.6^{+1.6}_{-2.6}$ & $8.5^{+4.7}_{-4.6}$  & $1.3^{+6.1}_{-1.3}$ & $0.008^{+0.005}_{-0.005}$ &  $0.992^{+0.005}_{-0.005}$ \\
\hline
$\pcb$ & & & & & & \\
$m_{\sigcp} + m_{\dvn}$   & $3.2^{+0.6}_{-0.9}$ & $14.9^{+1.2}_{-1.4}$ & $16.3^{+6.7}_{-7.4}$  & $4.3^{+9.2}_{-4.3}$ & $0.010^{+0.005}_{-0.004}$ &  $0.990^{+0.004}_{-0.005}$ \\
$m_{\sigcpp} + m_{\dvm}$  & $3.3^{+0.6}_{-0.9}$ & $15.6^{+1.0}_{-1.1}$ & $17.7^{+6.9}_{-8.2}$  & $2.9^{+8.3}_{-2.9}$ & $0.011^{+0.005}_{-0.005}$ &  $0.989^{+0.005}_{-0.005}$ \\
\hline
$\pcc$ & & & & & & \\
$m_{\sigcp} + m_{\dvn}$   & $1.5^{+0.7}_{-1.0}$ & $ 9.5^{+2.2}_{-5.1}$ & $3.5^{+4.2}_{-3.5}$  & $2.9^{+9.5}_{-2.9}$ & $0.002^{+0.003}_{-0.002}$ &  $0.998^{+0.002}_{-0.003}$ \\
$m_{\sigcpp} + m_{\dvm}$  & $1.8^{+0.6}_{-0.9}$ & $11.2^{+1.6}_{-2.5}$ & $5.4^{+4.2}_{-4.0}$  & $1.0^{+6.1}_{-1.0}$ & $0.003^{+0.003}_{-0.002}$ &  $0.997^{+0.002}_{-0.003}$ \\
\hline\hline
\end{tabular}
\end{scriptsize}
\caption{\label{tab.x1} Results obtained with $X=X_1+X_2=1$. The $\jpsi p$ and $\Sigma_c\bar{D}^{(*)}$ channels, 
which are labeled as $1$ and $2$ respectively, are included. } 
\end{table}

\begin{table}[htbp]
\centering
\begin{scriptsize}
\begin{tabular}{ c c c c c c c }
\hline\hline
Resonance & $|g_1|$ & $|g_2|$   & $\Gamma_1$ & $\Gamma_2$  & $X_1$   & $X_2$     \\
          &(GeV)  & (GeV)     & (MeV)      &  (MeV)      &         &     
\\ \hline
$\pca$ & & & & & & \\
$X=0.8$   & $2.3^{+0.7}_{-1.8}$ & $9.8^{+1.8}_{-2.5}$ & $7.1^{+5.0}_{-6.8}$  & $2.7^{+7.3}_{-2.7}$ & $0.007^{+0.005}_{-0.007}$ &  $0.793^{+0.007}_{-0.005}$ \\
$X=0.5$   & $2.4^{+0.7}_{-1.2}$ & $7.7^{+1.5}_{-2.0}$ & $8.1^{+5.1}_{-6.2}$  & $1.7^{+5.1}_{-1.7}$ & $0.008^{+0.005}_{-0.006}$ &  $0.492^{+0.006}_{-0.005}$ \\
\hline
$\pcb$ & & & & & & \\
$X=0.8$   & $3.2^{+0.7}_{-0.9}$ & $13.3^{+1.0}_{-1.3}$ & $17.2^{+7.6}_{-8.2}$  & $3.4^{+7.4}_{-3.4}$ & $0.011^{+0.005}_{-0.005}$ &  $0.789^{+0.005}_{-0.005}$ \\
$X=0.5$   & $3.4^{+0.7}_{-1.0}$ & $10.5^{+0.7}_{-1.0}$ & $18.5^{+9.0}_{-9.3}$  & $2.1^{+4.5}_{-2.1}$ & $0.012^{+0.006}_{-0.006}$ &  $0.488^{+0.006}_{-0.006}$ \\
\hline
$\pcc$ & & & & & & \\
$X=0.8$   & $1.6^{+0.7}_{-1.5}$ & $8.5^{+2.0}_{-4.5}$ & $4.1^{+4.6}_{-4.1}$  & $2.3^{+7.9}_{-2.3}$ & $0.002^{+0.003}_{-0.002}$ &  $0.798^{+0.003}_{-0.003}$ \\
$X=0.5$   & $1.7^{+0.8}_{-1.6}$ & $6.7^{+1.5}_{-3.3}$ & $5.0^{+5.1}_{-5.0}$  & $1.4^{+5.0}_{-1.4}$ & $0.003^{+0.003}_{-0.003}$ &  $0.497^{+0.003}_{-0.003}$ \\
\hline\hline
\end{tabular}
\end{scriptsize}
\caption{\label{tab.xsmall} Results obtained for $X=0.8$ and $X=0.5$ by including the $\jpsi p$ (labeled as 1) 
and $\Sigma_c \bar{D}^{(*)}$  (labeled as 2) channels. 
The values in the table are calculated by using the masses $\sigcp$ and $\bar{D}^{(*)0}$. } 
\end{table}

\begin{table}[htbp]
\centering
\begin{scriptsize}
\begin{tabular}{ c c c c c c c }
\hline\hline
Resonance & $|g_1|$ & $|g_2|$   & $\Gamma_1$ & $\Gamma_2$  & $X_1$   & $X_2$     \\
          &(GeV)  & (GeV)     & (MeV)      &  (MeV)      &         &     
\\ \hline
$\pca$ & & & & & & \\
$X=1.0$   & $4.0^{+2.0}_{-3.8}$ & $10.5^{+1.3}_{-2.5}$ & $6.8^{+5.4}_{-6.8}$  & $3.0^{+10.6}_{-3.0}$ & $0.09^{+0.16}_{-0.09}$ &  $0.91^{+0.09}_{-0.16}$ \\
$X=0.8$   & $4.2^{+2.0}_{-3.4}$ & $9.2^{+1.2}_{-2.0}$  & $7.5^{+5.5}_{-7.2}$  & $2.3^{+8.1}_{-2.3}$  & $0.10^{+0.16}_{-0.10}$ &  $0.70^{+0.10}_{-0.16}$ \\
$X=0.5$   & $4.5^{+2.0}_{-2.5}$ & $6.8^{+0.9}_{-1.2}$  & $8.5^{+5.7}_{-6.5}$  & $1.3^{+4.3}_{-1.3}$  & $0.11^{+0.17}_{-0.09}$ &  $0.39^{+0.09}_{-0.17}$ \\
\hline
$\pcb$ & & & & & & \\
$X=1.0$   & $3.8^{+0.7}_{-1.0}$ & $14.8^{+1.0}_{-1.3}$ & $16.4^{+6.8}_{-7.5}$  & $4.2^{+9.1}_{-4.2}$ & $0.03^{+0.01}_{-0.02}$ &  $0.97^{+0.02}_{-0.01}$ \\
$X=0.8$   & $3.9^{+0.8}_{-1.1}$ & $13.1^{+0.9}_{-1.1}$ & $17.3^{+7.7}_{-8.3}$  & $3.3^{+7.2}_{-3.3}$ & $0.03^{+0.01}_{-0.02}$ &  $0.77^{+0.02}_{-0.01}$ \\
$X=0.5$   & $4.0^{+1.0}_{-1.2}$ & $10.2^{+0.6}_{-0.8}$ & $18.6^{+9.2}_{-9.4}$  & $2.0^{+4.3}_{-2.0}$ & $0.03^{+0.02}_{-0.01}$ &  $0.47^{+0.01}_{-0.02}$ \\
\hline
$\pcc$ & & & & & & \\
$X=1.0$   & $1.7^{+0.9}_{-1.6}$ & $9.4^{+2.3}_{-5.0}$ & $3.5^{+3.7}_{-3.5}$  & $2.9^{+9.5}_{-2.9}$ & $0.005^{+0.007}_{-0.005}$ &  $0.995^{+0.005}_{-0.007}$ \\
$X=0.8$   & $1.9^{+0.8}_{-1.9}$ & $8.4^{+2.0}_{-4.4}$ & $4.1^{+4.6}_{-4.1}$  & $2.3^{+7.9}_{-2.3}$ & $0.006^{+0.008}_{-0.006}$ &  $0.794^{+0.006}_{-0.008}$ \\
$X=0.5$   & $2.0^{+0.9}_{-2.0}$ & $6.6^{+1.6}_{-3.2}$ & $5.0^{+5.1}_{-5.0}$  & $1.4^{+4.9}_{-1.4}$ & $0.008^{+0.008}_{-0.008}$ &  $0.492^{+0.008}_{-0.008}$ \\
\hline\hline
\end{tabular}
\end{scriptsize}
\caption{\label{tab.xlamcdv} Results obtained when including the $\Lambda_c^{+} \bar{D}^{*0}$ (labeled as 1) 
and $\Sigma_c^{+} \bar{D}^{*0}$ (labeled as 2) channels for $X=1.0$, $0.8$ and $0.5$.  } 
\end{table}

Summarizing, 
we have studied the newly discovered hidden-charm exotic states $P_c(4312)$, $P_c(4440)$ and $P_c(4457)$ by the LHCb Collaboration \cite{Aaij:2019vzc}.
We have first applied elastic effective-range expansion in the $\Sigma_c\bar{D}^{(*)}$ channel with the scattering length and the  effective range
fixed by reproducing the mass and width of every resonance separately. In all the cases one obtains values for these parameters of ${\cal O}(1)$~fm, which
supports their interpretation as composite resonances of $\Sigma_c\bar{D}^{(*)}$. We have also employed another coupled-channel approach involving the
two channels $\jpsi p$ and $\Sigma_c\bar{D}^{(*)}$ for each resonance, so that we require the saturation of the total width of the resonance.
By assuming some values for the total compositeness coefficients for these two channels, ranging from 0.5 to 1, we conclude that the weight of the
$\Sigma_c \bar{D}^{(*)}$ channel is much larger than the one for $\jpsi p$, in agreement with the ERE approach. We have also performed similar studies by including  alternatively the $\Lambda_c \bar{D}^{*}$ and $\Sigma_c\bar{D}^{(*)}$ as dynamical channels. The conclusions are basically the same as those obtained in the $\jpsi p$ and $\Sigma_c\bar{D}^{(*)}$ channels. 
Needless to say that more thorough
studies are needed, e.g., to disentangle the dynamics giving rise to the two nearby $P_c(4440)$ and $P_c(4457)$ resonances around the
$\sigcp\bar{D}^{*0}$ and $\sigcpp D^{*-}$ thresholds and its possible connection with isospin breaking (more likely in the case of composite resonances).

\section*{Acknowledgements}
This work is funded in part by the Natural Science Foundation of China under Grants No.~11575052, the Natural Science Foundation of Hebei Province under Contract No.~A2015205205, and the MINECO (Spain) and EU grant FPA2016-77313-P.


\begin{thebibliography}{90}

\bibitem{Aaij:2015tga} 
  R.~Aaij {\it et al.} [LHCb Collaboration],
  Phys.\ Rev.\ Lett.\  {\bf 115}, 072001 (2015)
  doi:10.1103/PhysRevLett.115.072001
  [arXiv:1507.03414 [hep-ex]].
 
\bibitem{ref.review}
See recent reviews: 
  Y.~R.~Liu, H.~X.~Chen, W.~Chen, X.~Liu and S.~L.~Zhu,
  arXiv:1903.11976 [hep-ph];

  H.~X.~Chen, W.~Chen, X.~Liu and S.~L.~Zhu,
  Phys.\ Rept.\  {\bf 639}, 1 (2016)
  doi:10.1016/j.physrep.2016.05.004
  [arXiv:1601.02092 [hep-ph]];

  F.~K.~Guo, C.~Hanhart, U.~G.~Mei{\ss}ner, Q.~Wang, Q.~Zhao and B.~S.~Zou,
  Rev.\ Mod.\ Phys.\  {\bf 90}, no. 1, 015004 (2018)
  doi:10.1103/RevModPhys.90.015004
  [arXiv:1705.00141 [hep-ph]];

  A.~Esposito, A.~Pilloni and A.~D.~Polosa,
  Phys.\ Rept.\  {\bf 668}, 1 (2016)
  doi:10.1016/j.physrep.2016.11.002
  [arXiv:1611.07920 [hep-ph]];
  
  A.~Ali, J.~S.~Lange and S.~Stone,
  Prog.\ Part.\ Nucl.\ Phys.\  {\bf 97}, 123 (2017)
  doi:10.1016/j.ppnp.2017.08.003
  [arXiv:1706.00610 [hep-ph]];

  R.~F.~Lebed, R.~E.~Mitchell and E.~S.~Swanson,
  Prog.\ Part.\ Nucl.\ Phys.\  {\bf 93}, 143 (2017)
  doi:10.1016/j.ppnp.2016.11.003
  [arXiv:1610.04528 [hep-ph]].

\bibitem{Aaij:2019vzc} 
  R.~Aaij {\it et al.} [LHCb Collaboration],
  arXiv:1904.03947 [hep-ex].



\bibitem{Chen:2019bip} 
  H.~X.~Chen, W.~Chen and S.~L.~Zhu,
  arXiv:1903.11001 [hep-ph].
  
\bibitem{Chen:2019asm} 
  R.~Chen, Z.~F.~Sun, X.~Liu and S.~L.~Zhu,
  arXiv:1903.11013 [hep-ph].

\bibitem{Guo:2019fdo} 
  F.~K.~Guo, H.~J.~Jing, U.~G.~Mei{\ss}ner and S.~Sakai,
  arXiv:1903.11503 [hep-ph].

  
\bibitem{Liu:2019tjn} 
  M.~Z.~Liu, Y.~W.~Pan, F.~Z.~Peng, M.~Sanchez Sanchez, L.~S.~Geng, A.~Hosaka and M.~P.~Valderrama,
  arXiv:1903.11560 [hep-ph].

\bibitem{He:2019ify} 
  J.~He,
  arXiv:1903.11872 [hep-ph].
  
\bibitem{Huang:2019jlf} 
  H.~Huang, J.~He and J.~Ping,
  arXiv:1904.00221 [hep-ph].
  
\bibitem{Xiao:2019mvs} 
  C.~J.~Xiao, Y.~Huang, Y.~B.~Dong, L.~S.~Geng and D.~Y.~Chen,
  arXiv:1904.00872 [hep-ph].
  
\bibitem{Shimizu:2019ptd} 
  Y.~Shimizu, Y.~Yamaguchi and M.~Harada,
  arXiv:1904.00587 [hep-ph].
  
\bibitem{Ali:2019npk} 
  A.~Ali and A.~Y.~Parkhomenko,
  arXiv:1904.00446 [hep-ph].
  
\bibitem{Cao:2019kst} 
  X.~Cao and J.~P.~Dai,
  arXiv:1904.06015 [hep-ph].
  
\bibitem{Gao:2018jhk} 
  R.~Gao, Z.~H.~Guo, X.~W.~Kang and J.~A.~Oller,
  arXiv:1812.07323 [hep-ph].

\bibitem{Kang:2016ezb} 
  X.~W.~Kang, Z.~H.~Guo and J.~A.~Oller,
  Phys.\ Rev.\ D {\bf 94}, no. 1, 014012 (2016)
  doi:10.1103/PhysRevD.94.014012
  [arXiv:1603.05546 [hep-ph]].

\bibitem{Kang:2016jxw} 
  X.~W.~Kang and J.~A.~Oller,
  Eur.\ Phys.\ J.\ C {\bf 77}, no. 6, 399 (2017)
  doi:10.1140/epjc/s10052-017-4961-z
  [arXiv:1612.08420 [hep-ph]].

  
\bibitem{Guo:2016wpy} 
  Z.~H.~Guo and J.~A.~Oller,
  Phys.\ Rev.\ D {\bf 93}, no. 5, 054014 (2016)
  doi:10.1103/PhysRevD.93.054014
  [arXiv:1601.00862 [hep-ph]].

\bibitem{Weinberg:1962hj} 
  S.~Weinberg,
  Phys.\ Rev.\  {\bf 130}, 776 (1963).
  doi:10.1103/PhysRev.130.776
  
\bibitem{Guo:2015daa} 
  Z.~H.~Guo and J.~A.~Oller,
  Phys.\ Rev.\ D {\bf 93}, no. 9, 096001 (2016)
  doi:10.1103/PhysRevD.93.096001
  [arXiv:1508.06400 [hep-ph]].
  
\bibitem{Oller:2017alp} 
  J.~A.~Oller,
  Annals Phys.\  {\bf 396}, 429 (2018)
  doi:10.1016/j.aop.2018.07.023
  [arXiv:1710.00991 [hep-ph]].
  
\bibitem{Meissner:2015mza} 
  U.~G.~Mei{\ss}ner and J.~A.~Oller,
  Phys.\ Lett.\ B {\bf 751}, 59 (2015)
  doi:10.1016/j.physletb.2015.10.015
  [arXiv:1507.07478 [hep-ph]].
  

  

\bibitem{Baru:2003qq} 
  V.~Baru, J.~Haidenbauer, C.~Hanhart, Y.~Kalashnikova and A.~E.~Kudryavtsev,
  Phys.\ Lett.\ B {\bf 586}, 53 (2004)
  doi:10.1016/j.physletb.2004.01.088
  [hep-ph/0308129].
  
\bibitem{Hanhart:2011jz} 
  C.~Hanhart, Y.~S.~Kalashnikova and A.~V.~Nefediev,
  Eur.\ Phys.\ J.\ A {\bf 47}, 101 (2011)
  doi:10.1140/epja/i2011-11101-9
  [arXiv:1106.1185 [hep-ph]].
  
\bibitem{Hyodo:2011qc} 
  T.~Hyodo, D.~Jido and A.~Hosaka,
  Phys.\ Rev.\ C {\bf 85}, 015201 (2012)
  doi:10.1103/PhysRevC.85.015201
  [arXiv:1108.5524 [nucl-th]].
  
\bibitem{Aceti:2012dd} 
  F.~Aceti and E.~Oset,
  Phys.\ Rev.\ D {\bf 86}, 014012 (2012)
  doi:10.1103/PhysRevD.86.014012
  [arXiv:1202.4607 [hep-ph]].
  
\bibitem{Sekihara:2014kya} 
  T.~Sekihara, T.~Hyodo and D.~Jido,
  PTEP {\bf 2015}, 063D04 (2015)
  doi:10.1093/ptep/ptv081
  [arXiv:1411.2308 [hep-ph]].
  
\bibitem{Fleming:2007rp} 
  S.~Fleming, M.~Kusunoki, T.~Mehen and U.~van Kolck,
  Phys.\ Rev.\ D {\bf 76}, 034006 (2007)
  doi:10.1103/PhysRevD.76.034006
  [hep-ph/0703168].
  
\bibitem{Valderrama:2012jv} 
  M.~P.~Valderrama,
  Phys.\ Rev.\ D {\bf 85}, 114037 (2012)
  doi:10.1103/PhysRevD.85.114037
  [arXiv:1204.2400 [hep-ph]].
  
\bibitem{Wu:2010jy} 
  J.~J.~Wu, R.~Molina, E.~Oset and B.~S.~Zou,
  Phys.\ Rev.\ Lett.\  {\bf 105}, 232001 (2010)
  doi:10.1103/PhysRevLett.105.232001
  [arXiv:1007.0573 [nucl-th]].
  
\bibitem{Wu:2010vk} 
  J.~J.~Wu, R.~Molina, E.~Oset and B.~S.~Zou,
  Phys.\ Rev.\ C {\bf 84}, 015202 (2011)
  doi:10.1103/PhysRevC.84.015202
  [arXiv:1011.2399 [nucl-th]].
  
\bibitem{Wu:2012md} 
  J.~J.~Wu, T.-S.~H.~Lee and B.~S.~Zou,
  Phys.\ Rev.\ C {\bf 85}, 044002 (2012)
  doi:10.1103/PhysRevC.85.044002
  [arXiv:1202.1036 [nucl-th]].

\bibitem{kolck.190419.1} U.~van Kolck,
  Nucl.\ Phys.\ A {\bf 645}, 273 (1999)
  doi:10.1016/S0375-9474(98)00612-5
  [nucl-th/9808007].

\bibitem{Oller:1998zr} 
  J.~A.~Oller and E.~Oset,
  Phys.\ Rev.\ D {\bf 60}, 074023 (1999)
  doi:10.1103/PhysRevD.60.074023
  [hep-ph/9809337].

\bibitem{Tanabashi:2018oca}
  M.~Tanabashi {\it et al.} [Particle Data Group],
  Phys.\ Rev.\ D {\bf 98}, no. 3, 030001 (2018).
  doi:10.1103/PhysRevD.98.030001


\bibitem{Shen:2016tzq} 
  C.~W.~Shen, F.~K.~Guo, J.~J.~Xie and B.~S.~Zou,
  Nucl.\ Phys.\ A {\bf 954}, 393 (2016)
  doi:10.1016/j.nuclphysa.2016.04.034
  [arXiv:1603.04672 [hep-ph]].
  
  
\bibitem{Lin:2017mtz} 
  Y.~H.~Lin, C.~W.~Shen, F.~K.~Guo and B.~S.~Zou,
  Phys.\ Rev.\ D {\bf 95}, no. 11, 114017 (2017)
  doi:10.1103/PhysRevD.95.114017
  [arXiv:1703.01045 [hep-ph]].
  
\end{thebibliography}
\end{document}